\documentclass[12pt]{iopart}
\usepackage{iopams}
\usepackage{setstack}
\usepackage[dvips]{epsfig,graphics,graphicx}
\usepackage{graphicx}
\usepackage{accents}
\usepackage{hyperref}
\usepackage{subfigure}
\eqnobysec
\def\m@th{\mathsurround=0pt}
\mathchardef\bracell="0365
\def\upbrall{$\m@th\bracell$}
\def\undertilde#1{\mathop{\vtop{\ialign{##\crcr
??? $\hfil\displaystyle{#1}\hfil$\crcr
???? \noalign
???? {\kern1.5pt\nointerlineskip}
???? \upbrall\crcr\noalign{\kern1pt
?? }}}}\limits}

\mathchardef\braceup"0371
\def\upbroll{$\m@th\braceup$}
\def\underhat#1{\mathop{\vtop{\ialign{##\crcr
??? $\hfil\displaystyle{\widehat{#1}}\hfil$\crcr
??? $\hfil\displaystyle{#1}\hfil$\crcr
???? \noalign
???? {\kern1.5pt\nointerlineskip}
???? \upbrall\crcr\noalign{\kern1pt
?? }}}}\limits}
\newcommand{\be}{\begin{equation}}
\newcommand{\ee}{\end{equation}}
\newcommand{\bee}{\begin{displaymath}}
\newcommand{\eee}{\end{displaymath}}
\newcommand{\bea}{\begin{eqnarray}}
\newcommand{\eea}{\end{eqnarray}}

\newcommand{\qiv}{$q$-$\mathrm{P}_{\mathrm{IV}}$}
\newcommand{\uIV}{$\mathrm{ud}$-$\mathrm{P}_{\mathrm{IV}}$}
\newcommand{\QIV}{$q$-$\mathrm{P}_{\mathrm{IV}}$}
\newcommand{\piv}{$\mathrm{P}_{\mathrm{IV}}$}
\newtheorem{prop}{Proposition} 
\begin{document}
\title{An ultradiscrete matrix version of the fourth Painlev\'e equation}
\author{Chris M Field$^\dag$ and Chris M Ormerod$^\ddag$}
\address{School of Mathematics and Statistics F07, The University of Sydney, Sydney, Australia}
\eads{$\dag$ cmfield@maths.usyd.edu.au ~~~~ $\ddag$ chriso@maths.usyd.edu.au}

\begin{abstract}

\noindent
We establish a matrix generalization of the ultradiscrete fourth Painlev\'e equation (\uIV).
Well-defined multicomponent systems that permit ultradiscretization are obtained using an approach that
relies on a group defined by constraints imposed by the requirement of a consistent evolution of the systems.
The ultradiscrete limit of these systems yields coupled multicomponent ultradiscrete systems that generalize
\uIV.  The dynamics, irreducibility, and integrability of the matrix valued ultradiscrete systems are studied.
\end{abstract}

    \section{Introduction}

Discrete Painlev\'e equations are difference equation analogs of classical Painlev\'e
equations \cite{Painleve:Original} and have been extensively studied recently (see the review article \cite{Gramani:Discrete}). The ultradiscrete Painlev\'e equations are discrete equations considered to be extended cellular automata
(they may also be considered as piecewise linear systems) that are derived by applying the ultradiscretization process \cite{Toki:CA} to discrete Painlev\'e equations. This process has been accepted as one that preserves integrability
\cite{Isojima:Shine}. Particular indicators of integrability in the ultradiscrete
setting include the existence of a Lax pair \cite{Quispel:PL}, an analog of singularity confinement \cite{Joshi:Sing},
and special solutions \cite{Orm:hyper}.

 
All of the preceding examples arising from ultradiscretization are one-component (that is, scalar) systems. 
Generalizations of integrable systems to associative algebras have been considered for many years
(see \cite{Olver:Associative}, and references therein). However, the general methods and results 
previously obtained are inapplicable in the ultradiscrete setting, due to the requirement of a subtraction
free setting. We present for the first time a matrix generalization of an ultradiscrete system.

The constraints related to the subtraction free setting and consistent evolution are 
studied in a group theoretic approach, in which one may also describe the nature of the irreducible subsystems. 
As an application of this method, 
we introduce a matrix version of the \uIV{} of \cite{Kajiw:fourth} which is derived by applying the 
ultradiscretization procedure to \QIV, 

\be\label{qP4}
\begin{array}{c c c}
f_0(qt)&=& a_0 a_1 f_1(t) \frac{1+ a_2 f_2(t) + a_2 a_0 f_2(t) f_0(t)}{1+ a_0 f_0(t) + a_0 a_1 f_0(t) f_1(t)}\\
f_1(qt)&=& a_1 a_2 f_2(t) \frac{1+ a_0 f_0(t) + a_0 a_1 f_0(t) f_1(t)}{1+ a_1 f_1(t) + a_1 a_2 f_1(t) f_2(t)}\\
f_2(qt)&=& a_1 a_2 f_0(t) \frac{1+ a_1 f_1(t) + a_1 a_2 f_1(t) f_2(t)}{1+ a_2 f_2(t) + a_0 a_2 f_0(t) f_2(t)}.
\end{array}
\ee
With the explicit form of the matrices derived, the new systems can be considered as coupled multicomponent generalizations.
It should be stressed that the approach of this paper gives {\it all} possible ultradiscretizable matrix valued versions of (\ref{qP4}).

The reason for choosing \qiv{} is that it has already been thoroughly and expertly investigated
in the scalar case (i.e., when $f_i$ and $a_i$ are scalar) \cite{Kajiw:fourth}. 
In \cite{Kajiw:fourth}, \qiv{} was shown  to admit the action of the affine Weyl group
of type $A_2^{(1)}$ as a group of B\"acklund transformations, to have classical solutions
expressible in terms of $q$-Hermite-Weber functions, to have rational solutions,
and its connection with the classification of Sakai \cite{Sakai_2001} was also investigated.
Furthermore, the ultradiscrete limit was taken in \cite{Kajiw:fourth}, and was shown to also admit affine Weyl group
representations.
As \qiv{ } is such a rich system, and has already been well-studied, this makes it
a perfect system for the application of our approach of ultradiscrete matrix generalization.

Before turning to the derivation of matrix \uIV, ultradiscretization should be 
introduced in more detail, so that the reason for certain constraints 
given later will be clear.

The process is a way of bringing a rational expression, $f$, in variables (or parameters) $a_1, \ldots, a_n$ to a new expression, $F$, in new ultradiscrete variables $A_1, \ldots , A_n$, that are related to the old variables via the relation $a_i = e^{A_i/\epsilon}$ and limiting process
\be
F(A_1, \ldots , A_n) = \lim_{\epsilon \to 0} \epsilon \log f(a_1, \ldots,a_n).
\ee
In general it is sufficient to make the following correspondences between binary operations 

\be\label{corresp}
\begin{array}{c c c}
a+ b &\to& \max(A,B)\\
a b &\to& A + B\\
a/b &\to& A-B .
\end{array}
\ee

This process is a way in which we may take an integrable mapping over the positive
real numbers $\mathbb{R}^+$ to an integrable mapping over the max-plus semiring \cite{Gram:CA}. 
The requirement that the pre-ultradiscrete equations are subtraction
free expressions of a definite sign is a more stringent restraint in the
matrix setting than the one-component setting, and it is this requirement which motivates the 
particular form of our matrix system.

The outline of this paper is as follows.  In section \ref{SecAA}, a \qiv{} is derived in the noncommutative 
setting, where the dependent variables take their values in an associative algebra.
In section \ref{SecMS} conditions on the matrix forms of the dependent variables and parameters of \qiv{} are 
derived such that it has a well-defined evolution and is ultradiscretizable. The group theoretic approach
is adopted to describe the constraints on the system. In section \ref{SecUD}
the ultradiscrete version of this system is derived, and some of the rich phenomenology of the derived matrix valued 
ultradiscrete \piv{} is displayed and analyzed in section \ref{SecPh}.

    \section{Symmetric \qiv{} on an associative algebra}\label{SecAA}

In this section it is shown that the symmetric \qiv{} of \cite{Kajiw:fourth} 
can be derived from a Lax formalism in the \emph{noncommutative} setting, where the
dependent variables $\{f_i\}$ take values in an a priori arbitrary associative algebra, $\mathcal{A}$,
with unit $I$ over a field $\mathbb{K}$ (when we turn to ultradiscretization,
the requirement of a field will be modified, 
but not in such a way as to affect the derivation from a Lax pair). 
This puts the present work in the context of other
recent work on integrable systems such as \cite{Mikhailov:Associaitve} and 
\cite{Olver:Associative} where the structure of integrable ODEs and PDEs (respectively)
was extended to the domain of associative algebras, and
\cite{Balandin:Associative} where Painlev\'e equations
were defined on an associative algebra (see also \cite{Olver:Associative}).
This trend has also been present in work on discrete integrable systems, such as
\cite{Bob:associative} where the higher dimensional
consistency (consistency around a cube) property
was investigated for integrable partial difference equations
defined on an associative algebra, and \cite{cmf:exactsolns} where an initial
value problem on the lattice KdV with dependent variables taking values in an associative algebra
was studied, leading to exact solutions. 

The auxiliary (spectral) parameter $x$, time variable $t$ and constant $q$
belong to the field $\mathbb{K}$.
The dependent variables $\{f_i\} \in \mathcal{A}$, 
system parameters $\{b_i\} \in \mathcal{A}$ ($i \in \{0,1,2\}$), and we define

    \be\label{Gam1}
        \Gamma_{i} := I + b^3_i f_i + b^3_i b^3_{i+1} f_i f_{i+1},
    \ee
up to an arbitrary ordering of the $b_j^3$ and $f_j$ factors.
(It will be shown that the ordering of these factors within $\Gamma_i$ is of no
consequence for either the integrability of the system or the existence
of a well defined evolution in the ultradiscrete limit.)
The invertibility of these expressions is assumed, that is $\Gamma_i^{-1} \in \mathcal{A}$.

We derive the system from a linear problem to settle other ordering issues in the 
noncommutative setting.
The $q$-type Lax formalism is given by

    \be
        \phi(qx,t) = L(x,t) \phi(x,t) \quad , \quad \phi(x,qt) = M(x,t) \phi(x,t)
    \ee
where
\numparts
    \bea
L(x,t) \ =\ \left( \begin{array}{ccc}
(1+x^2)I&b_0 f_0 t^{-\frac{2}{3}}&0\\0&(1+x^2)I&b_1 f_1 t^{-\frac{2}{3}} \\
b_2 f_2 t^{-\frac{2}{3}} &0&(1+x^2)I \end{array} \right) \ \\
M(x,t) \ =\ \left( \begin{array}{ccc}
0&b_2^{-2}\,\Gamma_2&0\\0&0&b_0^{-2}\,\Gamma_0\\b_1^{-2}\,\Gamma_1&0&0 \end{array} \right) . \label{qLIV}
    \eea
\endnumparts
The ultradiscrete version of this linear problem (for the usual commutative case) originally
appeared in \cite{Joshi:GeneralTheory}.

The compatibility condition for this linear problem reads

    \be
        M(qx,t) L(x,t) = L(x,qt) M(x,t),
    \ee
and leads to

    \be\label{qPIVpreq}
    \begin{array}{c c c}
     b_0 \,\overline{f}_0 &=& q^{2/3} \, b_2^{-2}\: \Gamma_2\, b_1 \, f_1 \,\Gamma_0^{-1} \, b_0^2,\\
     b_1 \,\overline{f}_1 &=& q^{2/3} \, b_0^{-2}\: \Gamma_0\, b_2 \, f_2 \,\Gamma_1^{-1} \, b_1^2, \\
     b_2 \,\overline{f}_2 &=& q^{2/3} \, b_1^{-2}\: \Gamma_1\, b_0 \, f_0 \,\Gamma_2^{-1} \, b_2^2,
    \end{array}
    \ee
where the overline denotes a time-update and $\overline{b}_i=b_i$.  

Following \cite{Kajiw:fourth}, we show a product of the dependent variables
can be regarded as the independent variable.
With $\{f^{-1}_i\} \in \mathcal{A}$, $\{b^{-1}_i\} \in \mathcal{A}$ (i.e., we are working with
a skew field) and specifying that the product $b_0 f_0 b_1 f_1 b_2 f_2$ is proportional to $I$, it is seen that  $b_0 f_0 b_1 f_1 b_2 f_2 = q c^2 I$ 
where $c \in \mathbb{K}$ and $\overline{c} = q c$.  Without loss of generality we set $c = t$. 
From now on 

    \be
        b_0 f_0 b_1 f_1 b_2 f_2 = q t^2 I
            \label{bfRES}
    \ee
will be imposed (so the algebra generated by all three $\{f_i\}$ and $I$ is not free). The invertibility of the algebra elements $\{f_i\}$ and $\{b_i\}$ is a consequence of the explicit matrix representation of these objects for the well-defined matrix systems studied in the next sections.

With the restriction 
    \be
        b_0^3 \, b_1^3\, b_2^3 = q I 
        \label{bbRES}
    \ee
imposed, the map (\ref{qPIVpreq}) is a noncommuting generalization of \QIV. If we specify that $b_i$ and
$f_i$ be matrix valued, the only requirement for a consistent evolution is that the $\Gamma_i$ are invertible.
This however is too general a system to be ultradiscretized, since in general we require the inverse to be
subtraction free.

If all variables commute, then after the change of variables $a_i := b_i^3$ the map reduces to \QIV, (\ref{qP4}), as presented in \cite{Kajiw:fourth}.

    \section{Ultradiscretizable matrix structure}\label{SecMS}

The conditions (\ref{bfRES}) and (\ref{bbRES}) can be used in conjunction with (\ref{qPIVpreq}) to define
constraints that lead to a consistent evolution on $\mathcal{A}$ as a free algebra with two constant (say $b_1$ and $b_2$)
and two variable (say $f_1$ and $f_2$) generators. Regarding these as 
$n \times n$ (or even infinite dimensional) matrices leads to multicomponent systems.  
However, the aim of the present work is to derive matrix
(or multicomponent) ultradiscrete systems, and hence, as we require the expressions to be subtraction free,
we have considerably less freedom than this general setting.

Due to this restriction, we restrict $\mathcal{A}$ to be the group of invertible non-negative matrices, that is
we set 
\be \label{subalg}
\mathcal{A} = S_n \ltimes \mathbb{K}^n
\ee
where $S_n$ is the symmetric group and $\mathbb{K}$ will further be restricted to be $\mathbb{R}^+$ in models
where we wish to perform ultradiscretization.
For our purposes $S_n$ is realized as $n \times n$ matrices of the form $\delta_{i \sigma(j)}$ for $\sigma \in S_n$.
(This group decomposition result can been seen in \cite{Bunini:Autos}.)
We define the homomorphism $\pi : \mathcal{A} \to \mathcal{A}/\mathbb{K}^n = S_n$ to be 
the homomorphism obtained as a result of the above semidirect product. This allows us to 
more easily deduce the form of the matrices $\{f_i\}$, $\{b_i\}$, that give a well-defined
evolution.

Since $\mathcal{A}$ is a semidirect product, the elements $b_i$ and $f_i$ 
can be uniquely written in the form

\be\label{bandf}
\begin{array}{c c c}
b_i&=& \beta_i \, s_i\\
f_i(t)&=& \digamma_i(t) \, z_i,  
\end{array}
\ee
where $\pi(b_i) = s_i \in S_n$, $\pi(f_i(t)) = z_i \in S_n$, and 
$\beta_i$ and $\digamma_i(t)$ are diagonal matrices containing 
the $n$ components of $b_i$ and $f_i(t)$ respectively (we leave the
matrix representation implicit). 

We now derive further restrictions on $\{s_i\}$ and $\{z_i\}$ such that
the evolution is consistent, and all terms in the map (such as the
$\Gamma_i$) remain in $\mathcal{A}$, (\ref{subalg}).

Consider the following form of $\Gamma_i$,

    \be
        \Gamma_{i} := I + b^3_i f_i + b^3_i b^3_{i+1} f_{i+1} f_{i}.
    \ee
As $\pi(I) = I$, $\pi(\Gamma_i) = I$ and this implies
\be\label{c1}
    s_i^3 z_i = I \qquad i \in \{ 0,1,2 \}.
\ee
This is the only condition that arises from the requirement that 
$\Gamma_i \in \mathcal{A}$, where $\mathcal{A}$ is given by (\ref{subalg}).
It is immediately seen that condition (\ref{c1}) is independent of the ordering of
the $b_i^3 f_i$ term in $\Gamma_i$.  
There are 24 possible orderings of $b_{i}^3 b_{i+1}^3 f_{i+1} f_{i}$
(we do not consider the possibility of splitting up the $b_i$ factors, as
$b_i^3 =: a_i$ is the parameter in the commutative case \cite{Kajiw:fourth}).
Of these 24 possibilities, 8 also require the commutativity of
$s_1^3$ and $s_{j}^3$ (equivalently $z_i$ and $z_{i+1}$).
It is shown in \ref{gpapp} that these additional commutativity
relations do not change the restrictions on $\{s_i\}$ and $\{z_i\}$.  
(That is, commutativity
of $s_i^3$ and $s_j^3$ is a consequence of the full set of relations.)

Requiring the preservation of (\ref{bandf}) as the variables evolve,
the projection of (\ref{qPIVpreq}) onto $S_n$, with (\ref{c1}), gives
\be\label{ca}
    s_i^4 = s_{i+1}^2 s_{i-1}^2 \qquad i \in \{0,1,2\}.
\ee   
The projection of the constraints (\ref{bfRES}) and (\ref{bbRES})
onto $S_n$, with (\ref{c1}), gives

    \be\label{cb}
        s_2^2 s_1^2 s_0^2 = I
    \ee
and
    \be\label{cc}
        s_0^3 s_1^3 s_2^3 = I
    \ee
respectively.

Therefore, to give a consistent evolution that permits ultradiscretization, $\{s_i\}$
are homomorphic images of the group generators of

    \be\label{group}
G = \langle g_0, g_1, g_2 \,|\, g_0^4=g_1^2 g_2^2,\, g_1^4 = g_2^2 g_0^2, \,g_2^4=g_0^2 g_1^2,\, g_2^2 g_1^2 g_0^2 = 1,\, g_0^3 g_1^3 g_2^3 = 1 \rangle
    \ee
in $S_n$; $\{ z_i \}$ are given by (\ref{c1}).  The group $G$ has order 108.
The order of the generators of $G$ is shown to be 18 in \ref{gpapp}.

    \section{Ultradiscretization}\label{SecUD}

We now consider the ultradiscretization of the matrix valued systems derived
in the previous section.
The components of the ultradiscretized systems belong to the 
max-plus semiring, $S$, which is the set 
$\mathbb{R}\cup \{-\infty \}$ adjoined with the binary operations of $\max$ and $+$ 
(often called tropical addition and tropical multiplication). 
To map the pre-ultradiscrete expression to the max-plus semiring, 
we may simply make the
correspondences (\ref{corresp}) on the level of the components.
(So $-\infty$ becomes the additive identity and $0$ becomes the multiplicative identity.)
By ultradiscretizing matrix operations, we arrive at the following 
definitions of matrix operations over $S$. 
If $A = (a_{ij})$ and $B = (b_{ij})$, then following \cite{Quispel:PL}, 
we define tropical matrix addition and multiplication, 
$\oplus$ and $\otimes$, by the equations
\begin{eqnarray*}
(A \oplus B)_{ij} := \max(a_{ij},b_{ij})\\
(A \otimes B)_{ij} := \max_k (a_{ik} + b_{kj})
\end{eqnarray*}
along with a scalar operation given by 
\[
(\lambda \otimes A)_{ij} := (\lambda + a_{ij})
\]
for all $\lambda \in S$.  
In the ultradiscrete limit $0$ is mapped to $-\infty$, and $1$ is mapped to $0$; hence
the identity matrix, $I$, is the matrix with $0$s 
along the diagonal and $-\infty$ in every other entry.
In the same way it is clear what happens to matrix realizations 
of members of $S_n$ in the ultradiscrete limit.
 
An ultradiscretized member of the group $\mathcal{A}$, (\ref{subalg}),
has a decomposition of the form
\bee
    D  = \Delta \otimes T
\eee
(cf. equation (\ref{bandf})) where $\Delta$ has $-\infty$ for all off-diagonal
entries and $T$ is an ultradiscretization of an element of $S_n$.
Its inverse is given by 
\bee
    D^{-1}  = T^{-1} \otimes \Delta^{-1},
\eee
where $(\Delta^{-1})_{ii} \equiv -(\Delta)_{ii}$ and all off-diagonal
entries are $-\infty$.

As well as the matrix map, the correspondence
also allows us to easily write the Lax pair over the semialgebra.
\numparts
    \bea
\mathcal{L}(X,T) \ =\ \left( \begin{array}{ccc}
\max(0,2X)\otimes I &B_0 \otimes  F_0 \otimes - \frac{2T}{3}&-\infty \\
-\infty& \max(0,2X) \otimes I &B_1 \otimes F_1 \otimes -\frac{2T}{3} \\
B_2 \otimes F_2 \otimes-\frac{2T}{3} &-\infty &\max(0,2X)\otimes I \end{array} \right) \ \\
 \ \mathcal{M}(X,T) \ =\ \left( \begin{array}{ccc}
-\infty&B_2^{-2}\,\otimes \,\Gamma_2&-\infty\\
-\infty&-\infty &B_0^{-2}\,\otimes \,\Gamma_0\\
B_1^{-2}\,\otimes\,\Gamma_1&-\infty&-\infty \end{array} \right) . \label{UDqLIV}
    \eea
\endnumparts
Where the ultradiscretization of $\Gamma_i$, as given in (\ref{Gam1}), is the matrix
\be\label{order}
\Gamma_i = I \oplus \left( B_i^3 \otimes F_i^3\right) \oplus \left( B_i^3 \otimes B_{i+1}^3 \otimes F_i\otimes F_{i+1}\right).
\ee
The compatibility condition reads
\be
        \mathcal{M}(X+Q,T) \otimes \mathcal{L}(X,T) = \mathcal{L}(X,T+Q)\otimes \mathcal{M}(X,T),
\ee
and gives the ultradiscrete equation over an associative $S$-algebra

\be\label{mudP4}
\begin{array}{c c c}
B_0 \otimes \overline{F}_0 &=& \frac{2}{3}Q \otimes B_2^{-2} \otimes \Gamma_2 \otimes B_1\otimes F_1 \otimes \Gamma_0^{-1} \otimes B_0^2, \\
B_1 \otimes \overline{F}_1 &=& \frac{2}{3}Q \otimes B_0^{-2} \otimes \Gamma_0 \otimes B_2\otimes F_2 \otimes \Gamma_1^{-1} \otimes B_1^2, \\
B_2 \otimes \overline{F}_2 &=& \frac{2}{3}Q \otimes B_1^{-2} \otimes \Gamma_1 \otimes B_0\otimes F_0 \otimes \Gamma_2^{-1} \otimes B_2^2.
\end{array}
\ee

The ultradiscrete version of the restrictions (\ref{bfRES}) and (\ref{bbRES}) are
\be
B_0\otimes F_0 \otimes B_1 \otimes F_1 \otimes B_2 \otimes F_2 = (Q + 2T)\otimes I
\ee
and
\be
B_0^3 \otimes B_1^3 \otimes B_2^3 = Q \otimes I .
\ee
(Of course, it would have been equally legitimate to apply the correspondence on the level
of the map (\ref{qPIVpreq}) without starting from a derivation from the ultradiscretized Lax pair.)

It is easily seen that if $2 Q/ 3$, the parameter
$T$, and all components of the map
belong to $\mathbb{Z}$ then at all time-steps all components
(not formally equal to $-\infty$) belong to $\mathbb{Z}$.
It is this property which motivates the term `extended cellular automata'.
 
    \section{Phenomenology}\label{SecPh}

As mentioned in the above discussion, we are required to find homomorphic images of the group $G$ in $S_n$. To do this, 
we use the computer algebra package Magma. The homomorphic images of $G$ in $S_n$ give rise to reducible and irreducible
subgroups, which in turn translate to reducible and irreducible matrix valued systems. 
By definition, the reducible systems
are decomposable into irreducible systems, and hence we restrict our attention to the irreducible cases. 

We may use any homomorphism to induce a group action of $G$ onto a set of $n$ objects. In this manner, we may state by the
orbit stabilizer theorem that the size of any orbit of $G$ must divide the order of the group. Since the group has order 108,
this implies the irreducible images of $G$ be of sizes that divide 108. In terms of matrix valued systems, the implication
is that any irreducible matrix valued systems are of sizes that divide 108. 

The lowest rank cases of the homomorphic images of the generators of $G$ in $S_n$ are given in table \ref{homomorphisms}
using the standard cycle notation for the symmetric group. 
The rank 1 case is well understood \cite{Kajiw:fourth}; hence we turn to the rank 2 case. For the examples presented here, we restrict 
our attention to the ordering within the $\{\Gamma_i \}$, (\ref{order}).

\begin{table}
\begin{center}
\begin{tabular}{ | c | c | c | c |} \hline
Rank & $g_0$ & $g_1$ & $g_2$ \\ \hline 
$1$ & $1$ & $1$ & $1$ \\\hline
$2$ & $(1,2)$ & $(1,2)$ & $1$ \\\hline 
$3$ & $(1,2,3)$ & $(1,2,3)$ & (1,2,3) \\
& $(1,2,3)$ & $(1,3,2)$ & $1$ \\ \hline
$4$ & $(1,2)(3,4)$ & $(1,3)(2,4)$ & $(1,4)(2,3)$ \\ \hline
$6$ & $(1,2,3,4,5,6)$ & $(1, 6, 5, 4, 3, 2)$ & $1$ \\
 & $(1, 2)(3, 4)(5, 6)$ & $(1, 3, 6)(2, 4, 5)$ & $(1, 5, 3, 2, 6, 4)$ \\
 & $(1,2,3)(4,5,6)$ & $(1, 4, 3, 6, 2, 5)$ & $(1, 4, 3, 6, 2, 5)$ \\\hline
\end{tabular}
\caption{\label{homomorphisms}Lowest rank cases of homomorphic images of the generators of $G$ in $S_n$.}
\end{center}
\end{table}

\begin{figure}[!ht]
\subfigure[$F_0$: both components]{\includegraphics[width=5.5cm]{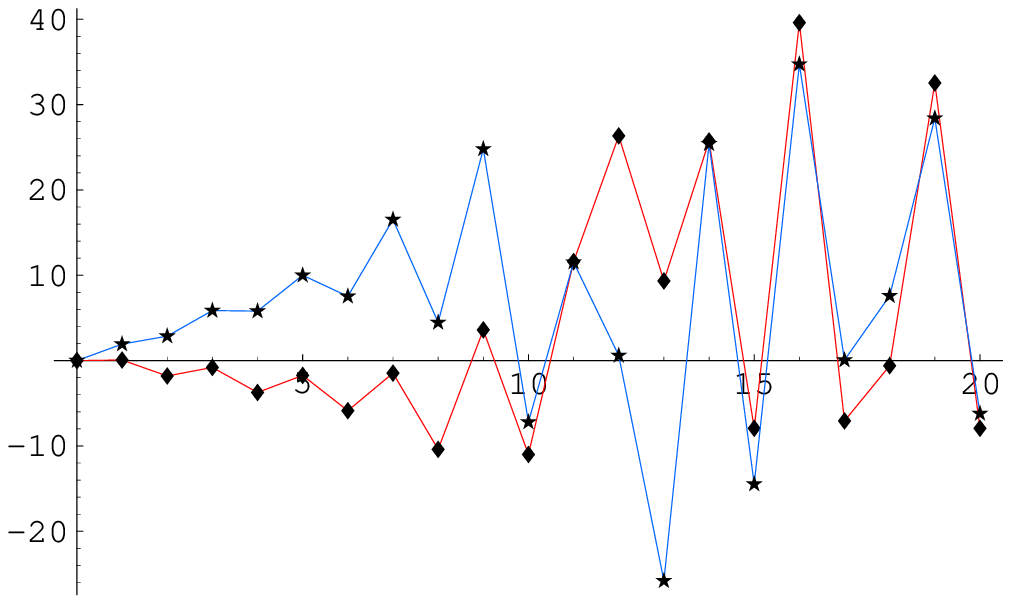}}
\subfigure[$F_1$: both components]{\includegraphics[width=5.5cm]{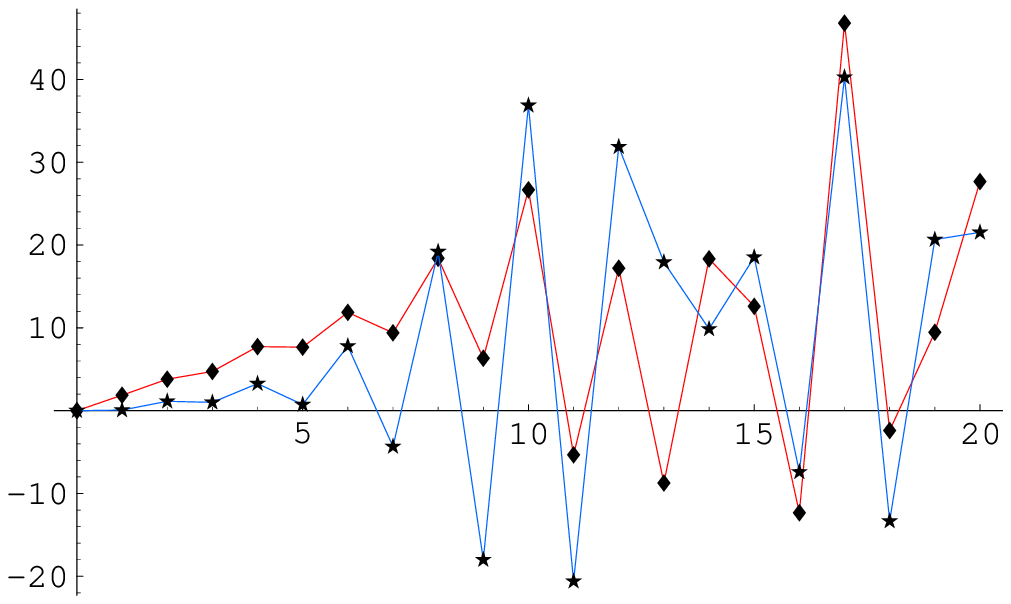}}
\subfigure[$F_2$: both components]{\includegraphics[width=5.5cm]{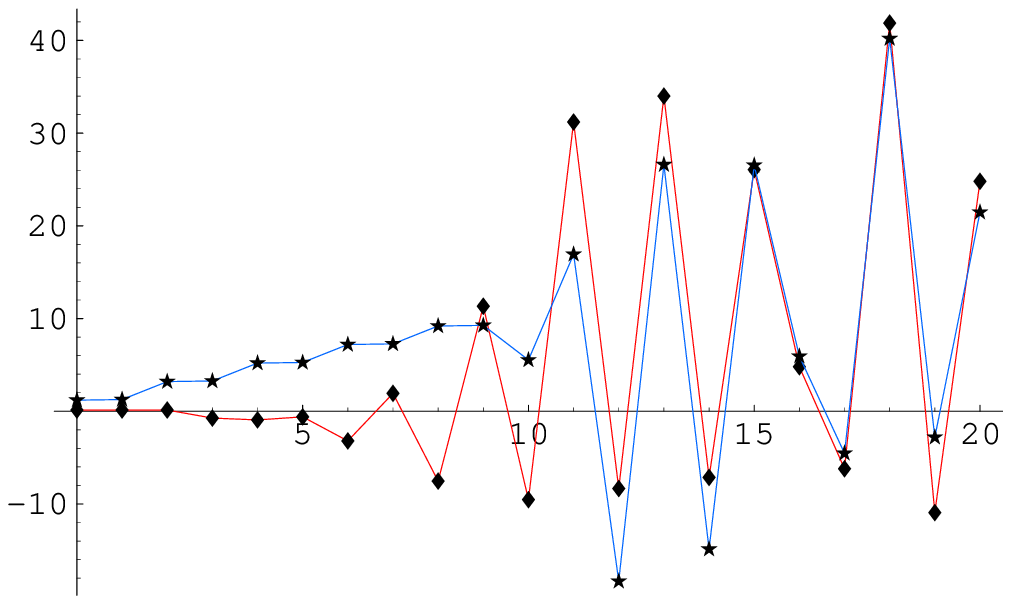}}
\caption{Generic behavior of the rank 2 case. Component values are plotted against time step values\label{H2figure}.}
\end{figure}

Typical behavior of the rank 2 map is shown in figure \ref{H2figure}. The initial conditions and 
parameter values in this case are
\[ 
B_0= \left( \begin{array}{ c c } -\infty & 0 \\ 0 & -\infty \end{array} \right) \qquad \, B_1= \left( \begin{array}{ c c } -\infty & 0 \\ \frac{4}{5} & -\infty \end{array} \right)\]
\[
F_0= \left( \begin{array}{ c c } -\infty & 0 \\ 0 & -\infty \end{array} \right) \qquad \, F_1= \left( \begin{array}{ c c } -\infty & 0 \\ 0 & -\infty \end{array} \right)
\]
where $B_2$ and $F_2$ are determined by the constraints, and $Q = 1$. For most initial conditions and
parameter values, the behavior has a similar level of visual complexity.

\begin{figure}[!ht]
\subfigure[$F_0$: both components]{\includegraphics[width=5.5cm]{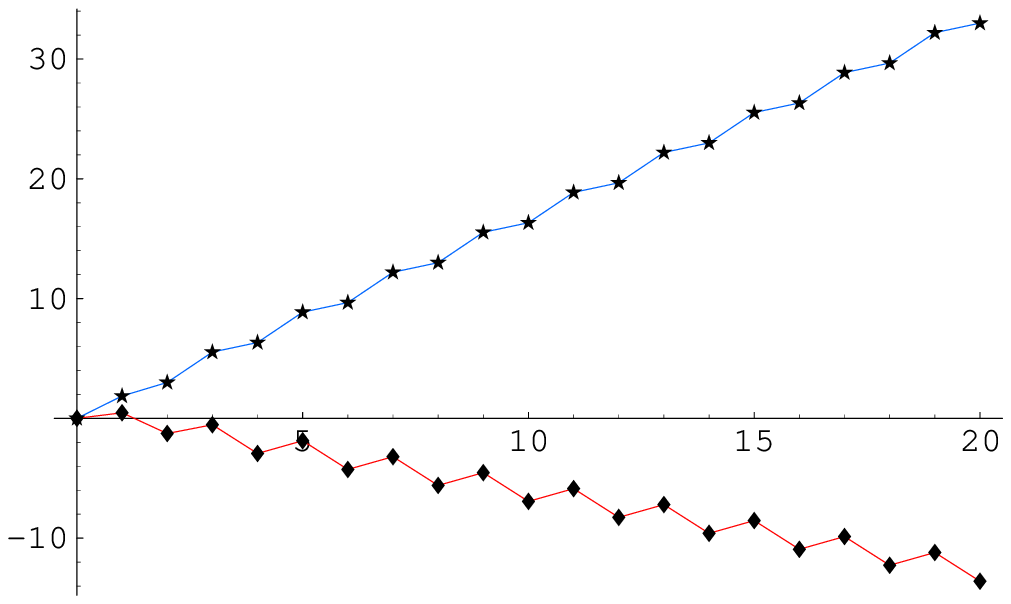}}
\subfigure[$F_1$: both components]{\includegraphics[width=5.5cm]{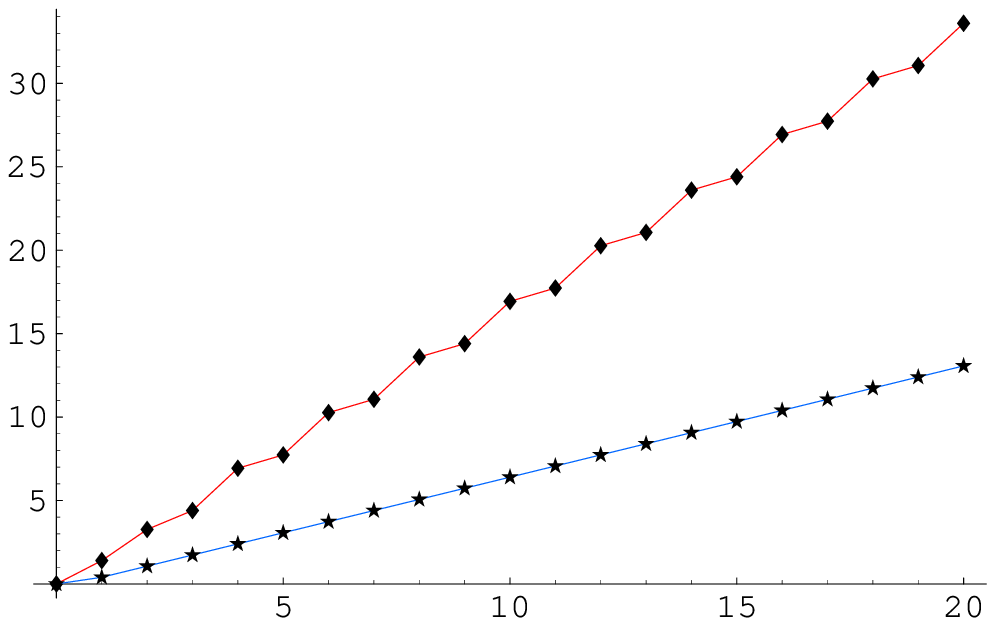}}
\subfigure[$F_2$: both components]{\includegraphics[width=5.5cm]{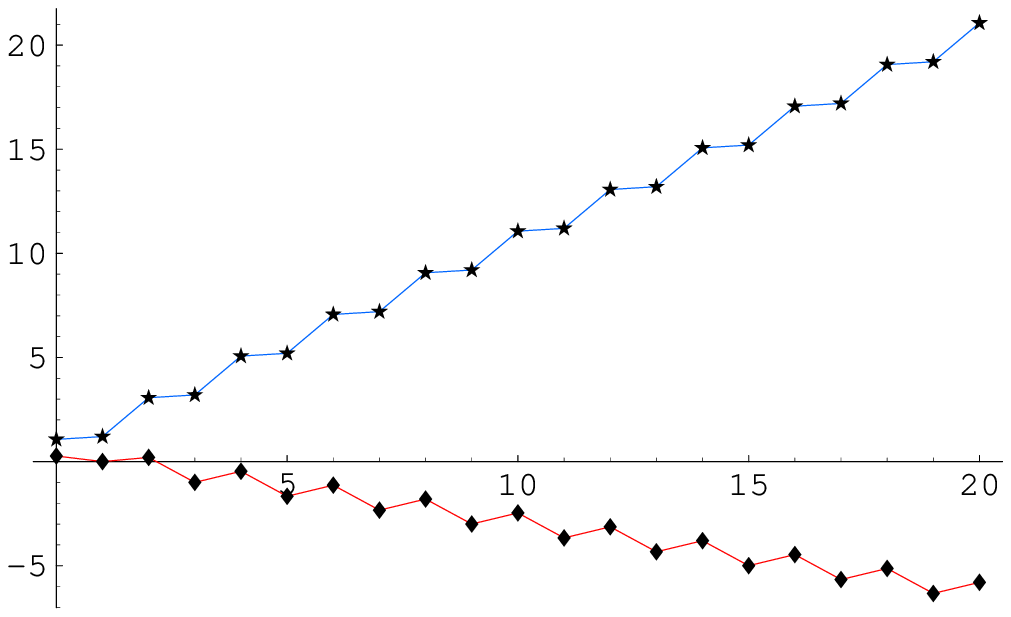}}
\caption{Some special behavior of the rank 2 case\label{Hyperfigure}.}
\end{figure}

It is a hallmark of the integrability of Painlev\'e systems that they possess special solutions such
as rational and hypergeometric functions \cite{Gramani:sp}. A remarkable discovery of our numerical investigations is that (\ref{mudP4})
displays special solution type behavior. These solutions only occur for specific parameter values and
initial conditions. One example of this comes at a surprisingly close set of parameters and initial 
conditions to those displayed by figure \ref{H2figure}. By setting the parameters to be
\[ 
B_0= \left( \begin{array}{ c c } -\infty & 0 \\ 0 & -\infty \end{array} \right) \qquad \, B_1= \left( \begin{array}{ c c } -\infty & 0 \\ \frac{3}{5} & -\infty \end{array} \right)\] 
with the same set of initial conditions, the behavior coalesces down to the much simpler form shown in
figure \ref{Hyperfigure}. 

The graphs of the single components in figure \ref{Hyperfigure} strongly resemble the recently discovered
ultradiscrete hypergeometric functions of \cite{Orm:hyper}.  This implies that the special solution
behavior shown here may be parameterized by a higher-dimensional generalization of the ultradiscrete
hypergeometric functions of \cite{Orm:hyper}.  
We discuss this possibility further in section \ref{Conc}.  
Behavior resembling rational solutions has also been observed in our computational investigations.

\begin{figure}[!ht]
\subfigure[$F_0$: all 3 components]{\includegraphics[width=5.5cm]{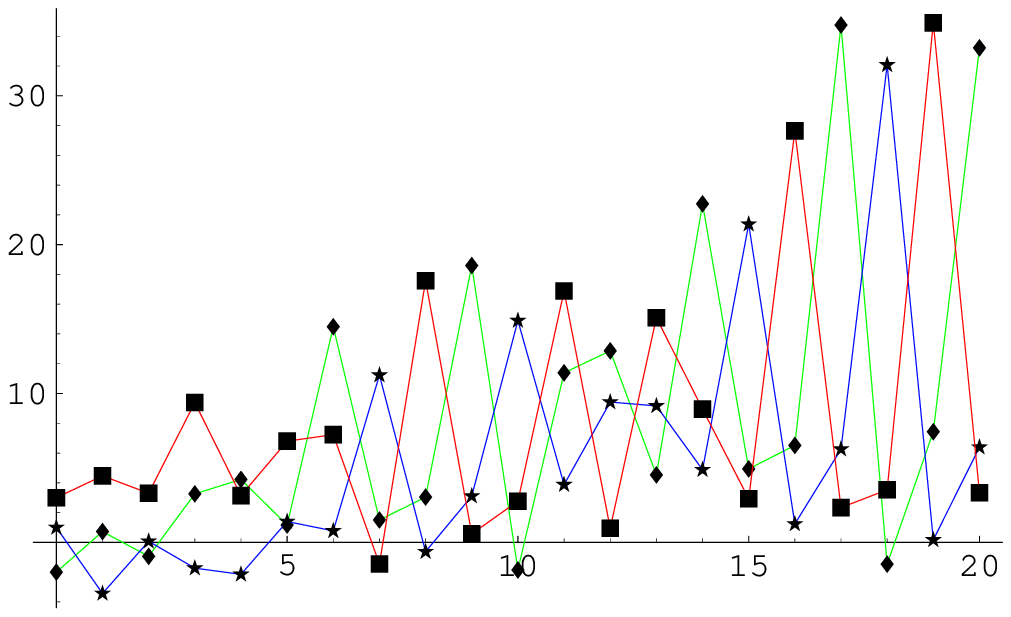}}
\subfigure[$F_1$: all 3 components]{\includegraphics[width=5.5cm]{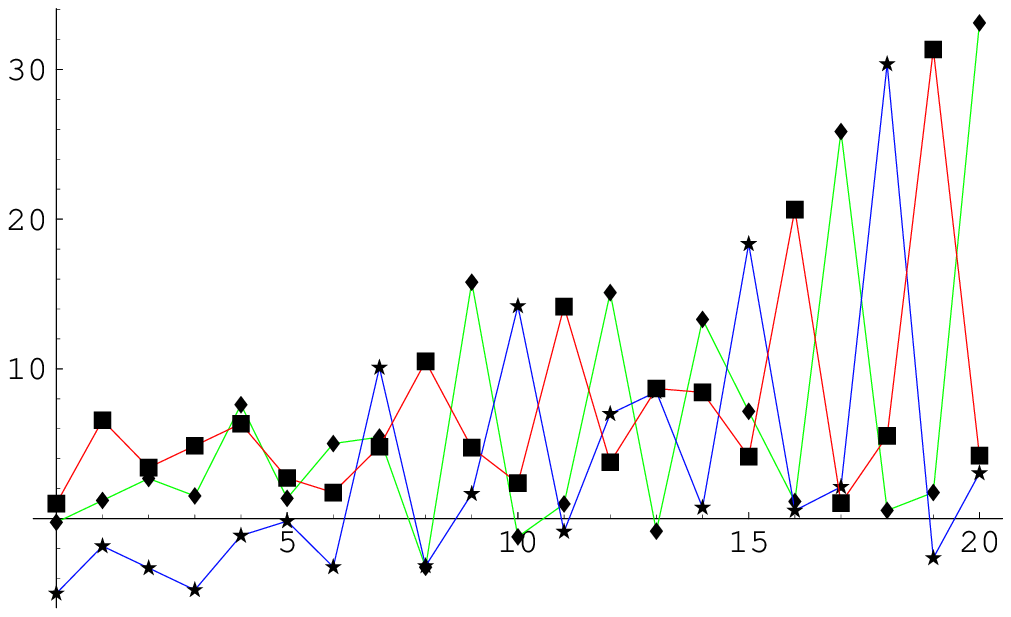}}
\subfigure[$F_2$: all 3 components]{\includegraphics[width=5.5cm]{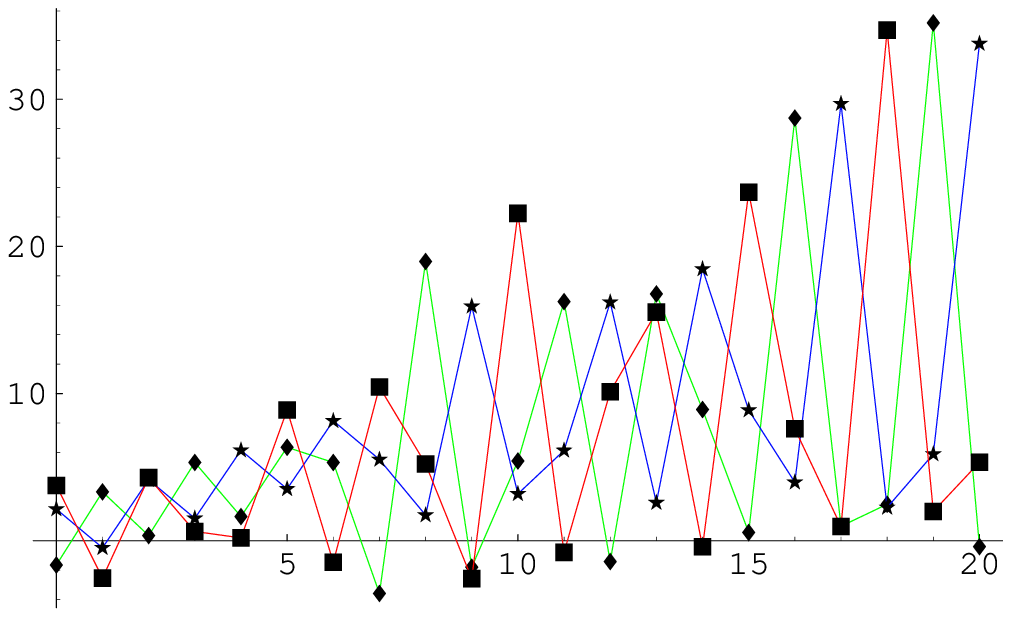}}
\caption{Generic behavior of the rank 3 case \label{H3figure}.}
\end{figure}

The typical behavior of the rank 3 map is shown in figure \ref{H3figure}. The initial conditions and 
parameter values are
\[ 
B_0= \left( \begin{array}{ c c c } -\infty & \frac{1}{5} & -\infty \\ -\infty & -\infty & \frac{1}{4} \\ -3 & -\infty & -\infty \end{array} \right) \, \qquad B_1=  \left( \begin{array}{ c c c } -\infty & -\infty & \frac{1}{7} \\ \frac{3}{5} & -\infty & -\infty \\ -\infty & -\frac{1}{2} & -\infty \end{array} \right)\]
\[
F_0=  \left( \begin{array}{ c c c } -2 & -\infty & -\infty \\ -\infty & 1 & -\infty \\ -\infty & -\infty & 3 \end{array} \right) \,  \qquad F_1= \left( \begin{array}{ c c c } -\frac{1}{4} & -\infty & -\infty \\ -\infty & -5 &-\infty \\ -\infty & -\infty & 1 \end{array} \right)
\]
where the coupling comes from the forms of the parameters. 

\begin{figure}[!ht]
\subfigure[$F_0$: all 3 components]{\includegraphics[width=5.5cm]{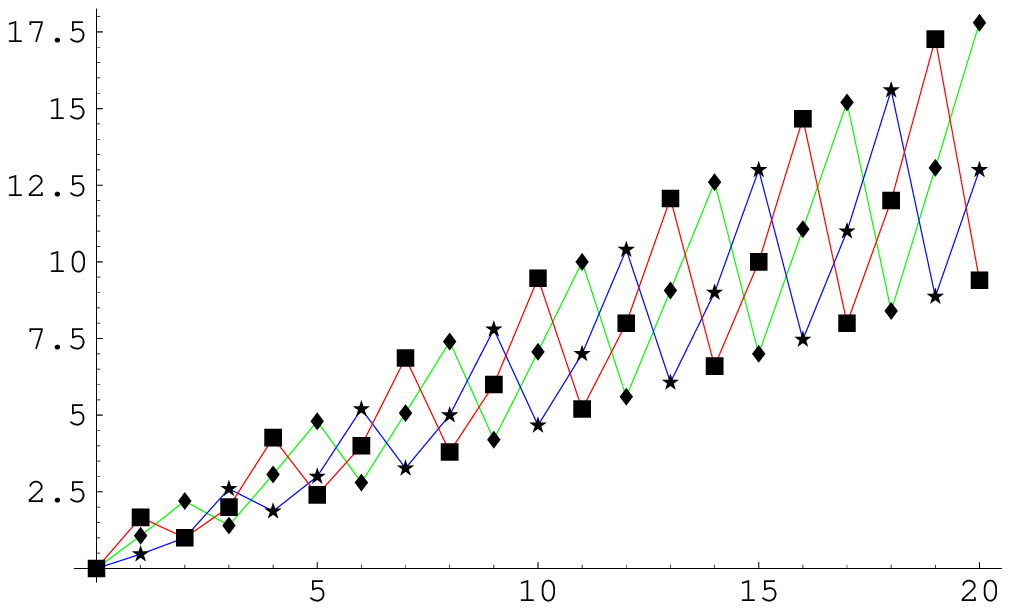}}
\subfigure[$F_1$: all 3 components]{\includegraphics[width=5.5cm]{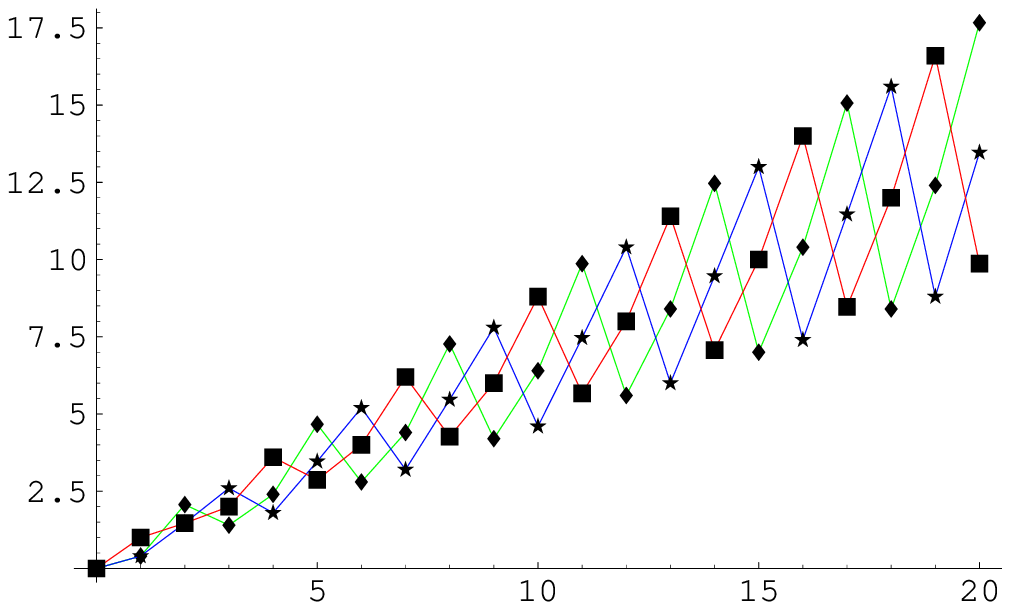}}
\subfigure[$F_2$: all 3 components]{\includegraphics[width=5.5cm]{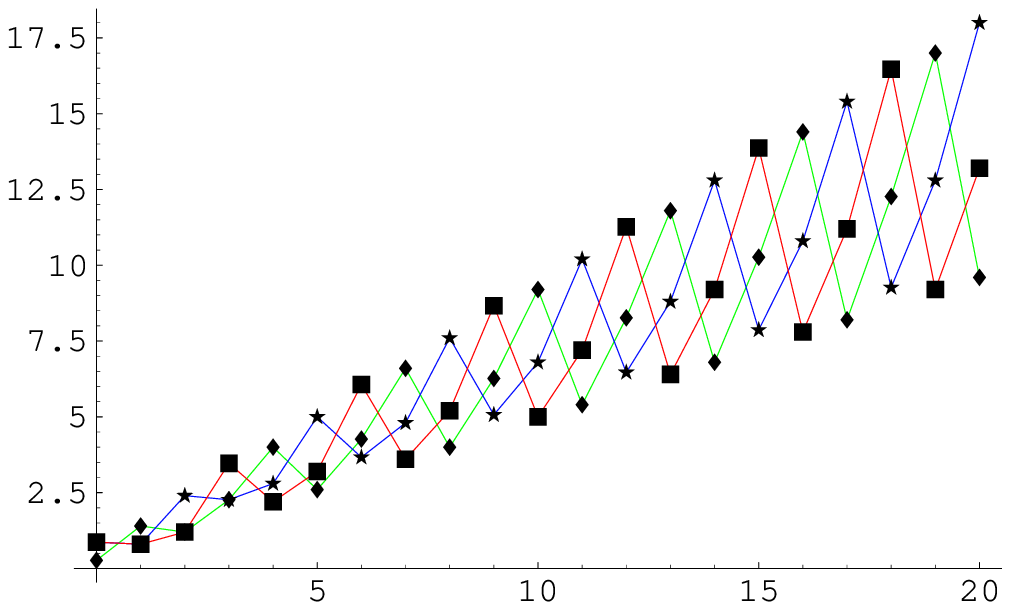}}
\caption{Some special behavior of the rank 3 case\label{H3Hyper} .}
\end{figure}

We also find behavior which we conjecture to be parameterized by higher-dimensional ultradiscrete hypergeometric 
functions. For initial conditions and parameters
\[ 
B_0= \left( \begin{array}{ c c c } -\infty & 0 & -\infty \\ -\infty & -\infty & 0 \\ 0 & -\infty & -\infty \end{array} \right) \, \qquad B_1= \left( \begin{array}{ c c c } -\infty & -\infty & 0 \\ \frac{3}{5} & -\infty & -\infty \\ -\infty & 0 & -\infty \end{array} \right)\]
\[
F_0= \left( \begin{array}{ c c c } 0 & -\infty & -\infty \\ -\infty & 0 & -\infty \\ -\infty & -\infty & 0 \end{array} \right) \, \qquad F_1= \left( \begin{array}{ c c c } 0 & -\infty & -\infty \\ -\infty & 0 &-\infty \\ -\infty & -\infty & 0 \end{array} \right)
\]
we obtain the behavior exhibited in figure \ref{H3Hyper}. 

    \section{Conclusions and discussion}\label{Conc}

We have presented a noncommutative generalization of \QIV. Conditions were derived
such that the matrix valued systems could be ultradiscretized. In section 4, the matrix 
generalization of ultradiscrete $\mathrm{P}_{IV}$ was presented. In section 5, a small snapshot
of the rich phenomenology was presented. 
Due to space restrictions, only certain aspects of this phenomenology 
was presented, yet our preliminary findings suggest many avenues for future research, including the generalization
of the results in \cite{Orm:hyper} to higher dimensional ultradiscrete hypergeometric functions. 

It is worth noting that a different generalization of \QIV, has been studied by Kajiwara et al. 
\cite{Kajiw:ddswWs}, \cite{Kajiw:qKP}. It would be interesting to know how both generalizations can be combined.

    \ack
The authors wish to thank Nalini Joshi for discussions on discrete Painlev\'e equations and ultradiscretization, and Stewart Wilcox for useful correspondence.  C.M. Field is supported by
the Australian Research Council Discovery Project Grant \#DP0664624. The research of C. Ormerod was supported in part by the
Australian Research Council grant \#DP0559019.

    \appendix 

    \section{Miscellaneous properties of the group $G$}\label{gpapp}

By deducing properties of the group $G$ presented in (\ref{group}), we may deduce properties of our
elements $\{s_i\}$ and $\{z_i\}$ since the $\{s_i\}$ must be homomorphic images of the generators of
$G$, while the $\{z_i\}$ are determined by the $\{ s_i \}$ via (\ref{c1}).

\begin{prop} \label{6}

\bea 
 g_0^6 = g_1^6 = g_2^6. \label{d1}
\eea
\end{prop} 
\textbf{Proof}
Constraint (\ref{ca}) implies 
\bee
    g_2^2 = g_1^{-2} g_0^4 = g_1^4 g_0^{-2}.
\eee
Therefore $g_0^6=g_1^6$, and similarly we have the full proof. $\square$

\noindent
(Note that this implies $[g_i,g_j^6]=0$.)

\begin{prop} Group elements $\{g_i\}$ have order 18.
\end{prop}
\textbf{Proof}
As $g_1^6 = g_0^6$ it follows from constraints (\ref{ca}) and (\ref{cb}) that
\bee
g_0^8 = g_2^2 g_0^2 g_2^{-2}.
\eee
Hence
\bee
g_0^{24} = g_2^2 g_0^6 g_2^{-2} = g_0^6
\eee
and therefore
\be \label{d2}
    g_0^{18} = I.
\ee
The proofs of 
\bee
g_1^{18} = I \qquad , \qquad g_2^{18} = I
\eee
proceed in the same manner. $\square$

\begin{prop}
\bea
\label{d3}    [g_i^{3},g_j^{3}] = 0.
\eea
\end{prop}
\textbf{Proof}
Using (\ref{d1}), equation (\ref{cc}) shows us that 
\bee
    g_0^6 = g_1^{-3} g_0^{-3} g_1^{-3} g_0^{-3},
\eee
further application of (\ref{d1}) reveals
\bee
    g_0^9 = g_0^{-6} g_1^{3} g_0^{-3} g_1^{-3},
\eee
hence, using (\ref{d2}),
\bee
    g_0^{-3} g_1^3 = g_1^3 g_0^{-3}.
\eee
Therefore we have the commutativity of $g_0^3$ and $g_1^3$, and
similarly we obtain (\ref{d3}). $\square$


\section*{References}


\begin{thebibliography}{10}

\bibitem{Balandin:Associative}
Balandin S P and Sokolov V V 1998
\newblock On the Painlev\'e test for non-abelian equations.
\newblock {\em Phys. Lett. A} {\bf 246} 267--272

\bibitem{Bob:associative}
Bobenko A I and Suris Yu B 2002
\newblock Integrable non-commutative equations on quad-graphs. The consistency approach.
\newblock {\em Lett. Math. Phys.} {\b 61}(3) 241--254


\bibitem{Bunini:Autos}
Bunina E I and Mikhal\"ev A V 2005
\newblock Automorphisms of the semigroup of invertible matrices with nonnegative elements.
\newblock {\it Fundam. Prikl. Mat.} {\bf 11} (2) 3--23. 

\bibitem{cmf:exactsolns}
Field C M, Nijhoff F W and Capel H W 2005
\newblock Exact solutions of quantum mappings from the lattice KdV as multi-dimensional operator difference equations.
\newblock {\em J. Phys. A} {\bf 38} 9503--9527


\bibitem{Gram:CA}
Grammaticos B, Ohta Y, Ramani A, Takahashi D and Tamizhmani K M 1997
\newblock Cellular automata and ultra-discrete Painlev\'e equations.
\newblock {\em Phys. Lett. A} {\bf 226} 53--58

\bibitem{Gramani:Discrete}
Grammaticos B and Ramani A 2004
\newblock Discrete Painlev\'e equations: a review. {\it Discrete integrable systems}
\newblock {\it Lecture Notes in Phys.} {\bf 644} Springer Berlin 245--321

\bibitem{Isojima:Shine}
Isojima S, Grammaticos B, Ramani A and Satsuma J 2006
\newblock Ultradiscretization without positivity
\newblock {\em J. Phys. A} {\bf 39} 3663–-3672

\bibitem{Joshi:LaxUD}
Joshi N, Nijhoff F W, and Ormerod C 2004
\newblock Lax pairs for ultra-discrete Painlev\'e cellular automata.
\newblock {\em J. Phys. A} {\bf 37} L559--L565

\bibitem{Joshi:Sing}
\newblock{Joshi N and Lafortune S 2005 How to detect integrability in cellular automata}\,
\newblock{{\it J. Phys. A}\, {\bf 38} (28)\, L499--L504}

\bibitem{Joshi:GeneralTheory}
Joshi N and Ormerod C 2007
\newblock The general theory of linear difference equations over the invertible max-plus algebra.
\newblock {\em Yet to appear}


\bibitem{Kajiw:fourth}
Kajiwara K, Noumi M, and Yamada Y 2001
\newblock A study on the fourth $q$-Painlev\'e equation.
\newblock {\em J. Phys. A} {\bf 34} 8563--8581

\bibitem{Kajiw:ddswWs}
Kajiwara K, Noumi M, and Yamada Y 2002
\newblock Discrete dynamical systems with $W(A^{(1)}_{m-1} \times A^{(1)}_{n-1})$ symmetry.
\newblock {\em Lett. Math. Phys.} {\bf 60} 211--219

\bibitem{Kajiw:qKP}
Kajiwara K, Noumi M, and Yamada Y 2002
\newblock $q$-Painlev\'e systems arising from $q$-KP hierarchy.
\newblock {\em Lett. Math. Phys.} {\bf 62} 259--268

\bibitem{Mikhailov:Associaitve}
Mikhailov A V and Sokolov V V 2000
\newblock Integrable odes on associative algebras.
\newblock {\em Comm. Math. Phys.} {\bf 211} 231--251

\bibitem{Olver:Associative}
Olver P J and Sokolov V V 1998
\newblock Integrable evolution equations on associative algebras.
\newblock {\em Comm. Math. Phys.} {\bf 193} 245--268

\bibitem{Orm:hyper}
Ormerod C 2006
\newblock Ultradiscrete hypergeometric solutions to Painlev\'e equations.
\newblock {\em Preprint} {\bf nlin.SI/0610048} 

\bibitem{Painleve:Original}
Painlev\'e P 1900
\newblock M\'emoire sur les \'equations diff\'erentielles dont l'int\'egrale g\'en\'erale est uniforme.
\newblock {\em Bull. Soc. Math. France} {\bf 28} 201--261

\bibitem{Gramani:Isomondromy}
Papageorgiou V G, Nijhoff F W, Grammaticos B and Ramani A 1992
\newblock Isomonodromic deformation problems for discrete analogues of Painlev\'e equations.
\newblock {\em Phys. Lett. A} {\bf 164}(1) 57--64 

\bibitem{Quispel:PL}
Quispel G R W, Capel H W, and Scully J 2001
\newblock Piecewise-linear soliton equations and piecewise-linear integrable maps.
\newblock {\em J. Phys. A} {\bf 34} 2491--2503

\bibitem{Gramani:Limit}
Ramani A, Grammaticos B and Hietarinta J 1991
\newblock Discrete versions of the Painlev\'e equations.
\newblock {\em Phys. Rev. Lett.} {\bf 67} 1829--1832

\bibitem{Gramani:sp}
Ramani A, Grammaticos B, Tamizhmani T, and Tamizhmani K M 2001
\newblock Special function solutions of the discrete Painlev\'e equations.
\newblock {\em Comp. and Math. with App.} {\bf 42} 603--614 

\bibitem{Gramani:Ultimate}
Ramani A, Takahashi D, Grammaticos B and Ohta Y 1998
\newblock{The ultimate discretisation of the Painlev\'e equations}\,
\newblock{{\it Phys. D}\, {\bf 114} (3-4)\, 185--196}

\bibitem{Sakai_2001}
Sakai H 2001
\newblock{Rational surfaces associated with affine root systems and geometry of the Painlev\'e equations}\, 
\newblock{{\it Comm. Math. Phys.}\, {\bf 220}\, 165--229}

\bibitem{Toki:CA}
Tokihiro T, Takahashi D, Matsukidaira J and Satsuma J 1996 
\newblock From soliton equations to integrable cellular automata through a limiting procedure.
\newblock {\em Phys. Rev. Lett.} {\bf 76} 3247--3250

\end{thebibliography}
\end{document}